\documentclass[pre,aps,twocolumn]{revtex4}

\usepackage{epsfig}
\usepackage{graphicx}
\usepackage{amssymb}

\newcommand{\be}{\begin{equation}}
\newcommand{\ee}{\end{equation}}
\newcommand{\bes}{\begin{eqnarray}}
\newcommand{\ees}{\end{eqnarray}}
\newcommand{\bma}{\left( \begin {array}}
\newcommand{\ema}{\end {array} \right)}

\newcommand{\um}{$\mu$m}
\newcommand{\ca}{$\text{Ca}^{2+}$}

\begin{document}
\title{Reaction-diffusion dynamics: confrontation between theory and
  experiment in a microfluidic reactor} 
\author{Charles N. Baroud$^\dag$\nocite{baroudaddress}, Fridolin Okkels, Laure M\'en\'etrier, Patrick Tabeling}
\affiliation{Laboratoire de physique statistique,
Ecole Normale Sup\'erieure, 24 Rue Lhomond, 75005 Paris\\ Ecole
Sup\'erieure de Physique et Chimie Industrielles, 8 Rue Vauquelin,
75005 Paris}
\date{\today} 


\begin{abstract}
We confront, quantitatively, the theoretical description of the
reaction-diffusion of a second order reaction to experiment. The
reaction at work is \ca/CaGreen, and the reactor is a T-shaped
microchannel, 10 $\mu$m deep, 200 $\mu$m wide, and 2 cm long.  The
experimental measurements are compared with the two-dimensional
numerical simulation of the reaction-diffusion equations.  We find
good agreement between theory and experiment.  From this study, one
may propose a method of measurement of various quantities, such as the
kinetic rate of the reaction, in conditions yet inaccessible to
conventional methods.
\end{abstract}

\maketitle

Chemical reaction processes taking place in spatially extended
reactors with no active mixing are described by reaction-diffusion
equations. In general, these equations are non linear and they possess
a rich variety of solutions, giving rise to many different
spatio-temporal patterns. Reaction-diffusion equations have been known
for more than a century but the first elementary solutions were
calculated much more recently by G\'alfi and R\'acz~\cite{galfi88}.
Their analytical solutions were obtained for a reaction of second
order in an infinite domain with stepwise initial conditions, in the
asymptotic limit of long times. Under these conditions, exact scaling
laws were derived showing, for instance, that the position of the
maximum of the production rate grows as $t^{1/2}$ while the width of
the reaction production zone grows as $t^{1/6}$. Furthermore, the
approach to the asymptotic state was studied by Taitelbaum et
al.~\cite{taitelbaum92} and they observed a slow approach to the
asymptotic state, which was modeled by using perturbation analysis and
could not be described by the long time asymptotics.

At the experimental level, one is often confronted with the additional
presence of advection in the fluid. For this reason, reactions are
often studied in small capillary tubes where no large scale flow is
present. Such systems have been used~\cite{koo91, park01} to study a
reaction process with initially separated reactants, validating the
power law predictions of G\'alfi and R\'acz at asymptotically long
times. Furthermore, recent experiments in porous media~\cite{leger99}
have shown very good agreement between theory and experiment in the
asymptotic regime, in the case of one diffusing reactant into a solid
substrate.

Here, we address the more general problem of two diffusing, initially
separated, reactants in a regime that spans both the initial
transients and the asymptotic state. In contrast with previously
published work~\cite{koo91,leger99}, we turn our attention to the
reaction {\it product} and its concentration, instead of the
production {\it rate}. This analysis takes into account the diffusion
of the product as well as the reagents. We compare numerical and
experimental measures of the reaction region, beyond the scaling of
the different quantities, to study the shape of the concentration
profiles.

Microfluidic circuits are used to achieve low Reynolds numbers
($Re\simeq1$), thus removing any effects of advection without
resorting to gels or other media. We believe that the detailed
comparison of the shape of the concentration profiles between
simulations and experiments allows one to assess to what extent, from
the dynamical viewpoint, modeling a reaction by a kinetics of a
prescribed order is accurate. This assessment is useful in view of
modeling more complex behaviors in laminar, chaotic, or turbulent
flows. In this regard, microsystems offer the opportunity to
accurately compare theory and experiment since simple geometries can
be achieved and, owing to the low Reynolds number at hand, the
hydrodynamics is well controlled. Diffusion is the only transport
mechanism leading the reagents to react together.

The channel geometry we consider here precisely allows the study of a
reaction-diffusion process with stepwise initial conditions.  The
micro-reactor is T-shaped, similar to that used in
Ref.~\cite{kamholz99}. In contrast with these authors, we use shallow
channels which favor two-dimensionality as discussed below. We will
further infer from the present work a practical method for measuring
the kinetics of chemical reactions in the sub-millisecond range, a
regime that is out of the reach of traditional stopped-flow
techniques~\cite{cussler}.

\begin{figure}[hb]
\includegraphics[width=8cm]{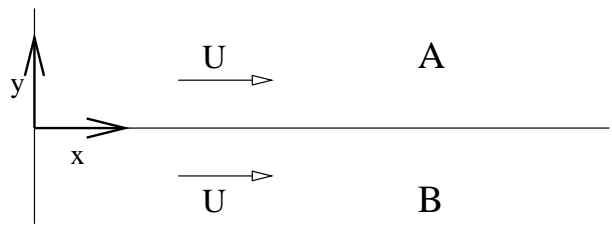}
\caption{Geometry of the theoretical problem.\label{scheme}}
\end{figure}

{\it Theoretical background}: Let us consider a second order 
irreversible reaction of the type $A+B \rightarrow C$, taking place 
in a two-dimensional space, within a flow driven at uniform velocity 
U, as shown in Fig.~\ref{scheme}. 

In the stationary state, after neglecting the derivative along $X$ in
the Laplacian operator, the reaction-diffusion equations read :

\bes
U \frac{\partial A}{\partial X} &=&D_a \frac{\partial^2 A}{\partial
Y^2} - kAB \nonumber \\
U \frac{\partial B}{\partial X} &=& D_b \frac{\partial^2 B}{\partial 
Y^2} - kAB
\label{eq:rd}
\ees

\noindent where $A$ and $B$ are the concentrations of the two
reactants A and B, $D_a$ and $D_b$ are the corresponding diffusion
coefficients, and $k$ is the chemical rate constant.  These equations
are identical to the typical reaction-diffusion equations, if we
replace variable $X$ by $Ut$, where $t$ is the time. This system must
be supplemented by an equation governing the evolution of product C,
given by :

\be
U \frac{\partial C}{\partial X}=D_c \frac{\partial^2 C}{\partial 
 Y^2} + kAB 
\label{eq:rp}
\ee

\noindent where $C$ is the concentration of product C, assumed inert,
and $D_c$ is the corresponding diffusion coefficient. As initial
conditions, we take the reagents to be separated with constant
densities for both $Y<0$ ($A=A_{0}$ and $B=0$) and $Y>0$ ($A=0$ and
$B=B_{0}$). We thus have here a six dimensional parameter space which
consists of the three diffusion coefficients of the two reagents and
the product, the initial concentration levels of the reagents, and the
reaction rate coefficient.

The system of Eqs.~\ref{eq:rd} can be non-dimensionalized by
introducing a characteristic length scale $\ell$, similar to the one
introduced in Ref.~\cite{galfi88}

\be
\ell^2 = \frac{\sqrt{D_a D_b}}{k\sqrt{A_{0} B_{0}}}.
\label{eq:ell}
\ee

\noindent This length is used to non-dimensionalize $X$ and $Y$ as:
$$ x=\frac{XD}{{\ell^2}U},\quad {y}=\frac{Y}{\ell}$$ where
$D=\sqrt{D_aD_b}$.  Using these new dimensionless variables and
introducing the dimensionless concentrations
$$a=\frac{A}{A_{0}}\quad \text{and} \quad b=\frac{B}{B_{0}},$$ the
governing equations for the problem become:

\bes
\frac{\partial a}{\partial {x}} &=&\chi \frac{\partial^2 a}{\partial
{y}^2} - \frac{a b}{\beta} \nonumber \\
\frac{\partial b}{\partial {x}} &=&\frac{1}{\chi}
\frac{\partial^2 b}{\partial {y}^2} - \beta ab 
\label{eq:rd-nd}
\ees

\noindent in which we define

\be
\chi=\sqrt{D_a/D_b},~~\beta=\sqrt{A_{0}/B_{0}}.
\ee

This system must also be supplemented by an equation prescribing the
evolution of the product concentration. Although the problem can be
solved in the general case, we restrict ourselves to the particular
case where the third diffusion coefficient, $D_c$, is equal to $D_a$
(assumed to be the smaller than $D_{b}$). The equation for the
evolution of the product concentration thus reads, in dimensionless
form:

\be
\frac{\partial c}{\partial {x}} = \frac{1}{\chi} \frac{\partial^2 c}{\partial {y}^2} + ab
\label{eq:rc-nd}
\ee

\noindent with $c=C/\sqrt{A_{0}B_{0}}$. Note that the reaction rate has
disappeared in Eqs.~\ref{eq:rd-nd} and~\ref{eq:rc-nd}. This means that
chemical kinetics does not play any dynamical role: changing $k$
rescales the variables of the problem but does not change the
structure of the solution. A consequence is that the number of
parameters is reduced to two, meaning that the full set of solutions
can be represented in a two-dimensional parameter space. A knowledge
of both parameters $\chi$ and $\beta$ is necessary to compare
simulations and experiments, if one is interested in both the
transient and asymptotic regimes. 

\begin{figure}
\includegraphics[width=9cm,height=5cm]{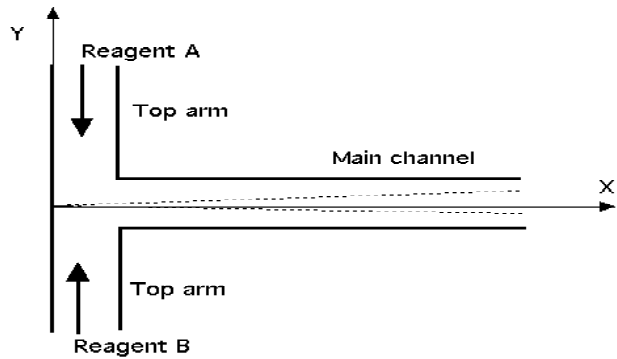}
\caption{geometry of the microreactor.\label{tchannel}}
\end{figure}

{\it Description of the experiment:} In our experiments, we used a
T-shaped microreactor, whose geometry is represented in
Fig~\ref{tchannel}. This geometry is similar to the one introduced in
Ref.~\cite{kamholz99}.

\begin{figure}
\includegraphics[width=9cm]{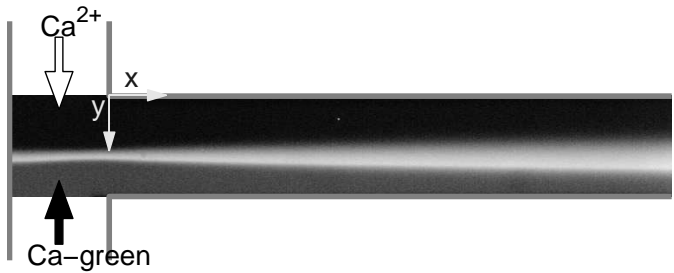}
\caption{Fluorescent microscopy image of the binding of \ca~ions with
Calcium-green marker in a T-channel. The channel width is 200~\um, the
depth is 10~\um. Here, the velocity in the main channel is
$U=0.083$~cm/s.\label{fig:experiment}}
\end{figure}

The channels were chemically etched in a glass wafer, which was then
anodically bonded to a silicon wafer to create a rectangular
cross-section. Throughout the experiments discussed here, the channel
depth was 10 $\mu$m and the width was 200 $\mu$m. The lengths of the T
arms were 1 cm while the main channel was 2 cm long. Two reactant
species are introduced through the two top branches of the T and they
react as they flow side by side in the main channel. We study the
reaction of Calcium ions as reactant A and Calcium Green (CaGreen)
from Molecular Probes~\cite{probes} as reactant B (as used, e.g.
in~\cite{monson00}). \ca~ions bind with the tracer Calcium Green,
significantly increasing its fluorescence~\cite{reaction}.
Epifluorescence microscopy is used to measure the fluorescence
emission, yielding spatial images such as Fig.~\ref{fig:experiment}.
The flows are driven by syringe pumps which also provide measurements
of the flow-rates in each separate branch; those ranged between 30 and
100~nl/mn. Owing to the aspect ratio of the micro-channel, the
velocity profile is parabolic across the channel depth and is
essentially uniform along $Y$ direction. Furthermore, the low Reynolds
number ensures that the fluid flow is invariant in the $x$-direction
after some entrance length on the order of the channel thickness.

In the conditions in which we operated (i.e. low flow-rates and thin
channels) the tracer concentration can be considered as uniform across
the channel depth. As a consequence, the concentration fields can be
treated as two-dimensional and Eqs.~\ref{eq:rd-nd} and~\ref{eq:rc-nd}
can be used to describe the system. We tested the two-dimensionality
property by performing a series of diffusion experiments (with no
reaction) of fluorescein in water. A fluorescein solution and pure
water are driven from the two top arms respectively. In the main
channel, they form a diffusing interface which progressively broadens
downstream. In thick channels, the broadening was found to display the
three-dimensional effects discussed in Refs.~\cite{kamholz99}
and~\cite{ismagilov00}, where the diffusion region scales as
$x^{1/3}$. In the 10~$\mu$m channels, 2D diffusion theory to agrees
well with experiment, with a scaling of the diffusion width as
$x^{1/2}$.

The comparison between the 2D theoretical model and the experiment is
further justified by the following reasoning: one may consider that
the reacting zone is homogeneous across the channel depth in the major
part of the region we analyze, for sufficiently thin channels. This
argument relies on estimations of the diffusive time across the
channel thickness and on the measurements with fluorescein discussed
above. We are thus working mostly in the so-called Taylor-Aris regime
where the parabolic velocity profile enhances the apparent diffusion
coefficient in the flow direction~\cite{pagitsas86}. One may show
however that the streamwise diffusion term is still negligible
compared to the transverse term, since the streamwise gradients are
very weak. These two features justify the use of the model of
Eqs.~\ref{eq:rd-nd} to compare with our experiment.

The diffusion coefficient of Calcium ions is equal to
$D_{\text{Ca}^{2+}}=7.9\times 10^{-6}~$cm$^2$/s~\cite{cussler}. The
diffusion coefficient of the CaGreen was unknown so we measured it
using the method described above, replacing the fluorescein solution
with a Calcium Green solution: We checked that the width of the
diffused zone increases as the downstream distance raised to the power
1/2, as expected from two-dimensional theory.  By fitting the
measurements of the width, we obtained an estimate for the Calcium
Green diffusion coefficient equal to $3.2\pm 0.3\times
10^{-6}$~cm$^{2}$/s.
 
Throughout the experiments, we kept the concentration of \ca ~constant
at [\ca]=1~mM, while the concentration of CaGreen was varied in the
range $0.1<\text{[CaGreen]}<0.5$~mM. Therefore, the value of $\beta$
varied between $1.4$ and $3.2$, while $\chi$ was fixed at $\chi=1.57$.

{\it Comparison between theory and experiment}: A typical
concentration field is shown in Fig~\ref{fig:experiment}. The black
region corresponds to the Calcium solution, the grey one to the
CaGreen solution, and the brighter region in the center to the bound
Ca-CaGreen complex. The background fluorescence, visible in the lower
part of the channel, is due to the fluorescence of unbound CaGreen.
The asymmetry in the reaction zone is due to a different rate of
diffusion of the reagents, because of their different initial
concentrations and different diffusion coefficients. This can be
physically explained by noting that the species with the higher
diffusion coefficient will diffuse further; furthermore the species
with the larger initial concentration will diffuse faster due to the
large gradient it experiences. In other words, $\beta$ and $\chi$ both
determine the shape of the reaction zone.

We were not able to obtain reliable information on the kinetic
coefficient $k$ of the \ca-Calcium Green reaction. We thus took $k$ as
a free fitting parameter, and adjusted it so as to optimize the
agreement between theory and experiment. The best fits were obtained
for $k=1.0\pm 0.47\times 10^6$~dm$^3$/(mol s), or a length scale
$\ell=0.84$~\um~for $\beta=\sqrt{2}$ and $\ell=1.0$~\um~for
$\beta=2$. These values of $k$ and $\ell$ are used throughout this
letter in comparing the experiments and numerics.

\begin{figure}[ht]
\includegraphics[width=9cm]{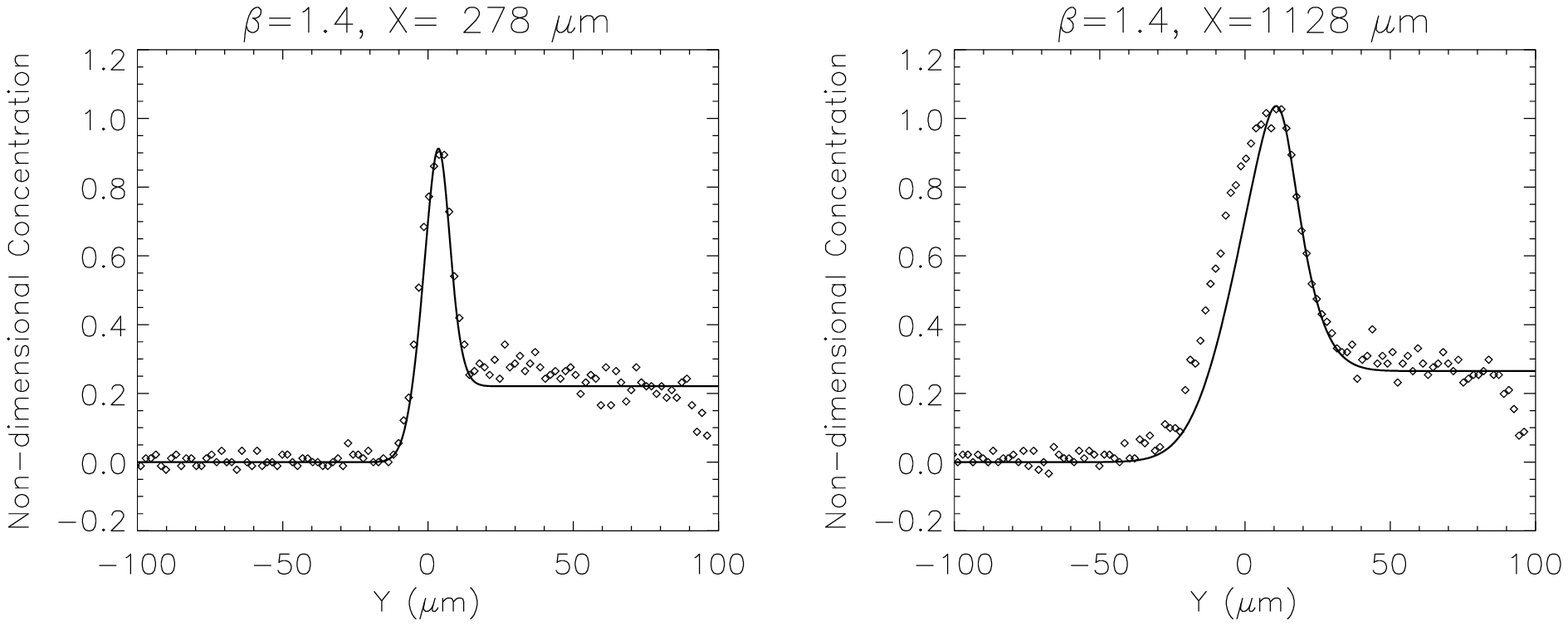}
\includegraphics[width=9cm]{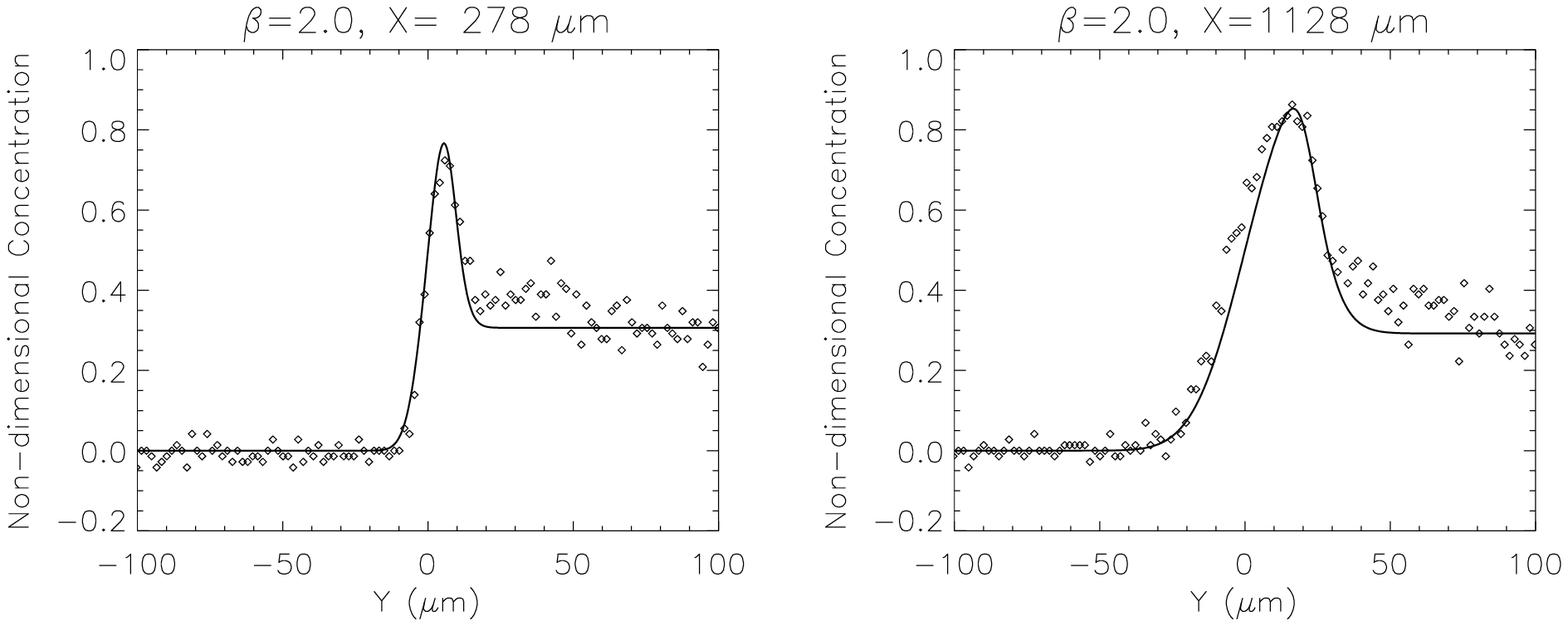}
\caption{Experimental measurements of the fluorescent intensity (dots)
compared with numerical solutions of $c$ (solid lines) at different
downstream locations for $\beta=\sqrt{2}$ and $2$. $U=0.334$~cm/s for
both cases.}
\label{fig:experiment-numerics1}
\end{figure}

Figure~\ref{fig:experiment-numerics1} represents the concentration
profiles of the Ca-CaGreen complex (dots) measured at various
locations downstream, for $\beta=\sqrt{2}$ and 2. The two locations we
consider here are $X=278$ and 1128~$\mu$m (for $\beta=\sqrt{2}$ these
correspond to $x=32$ and $216$; for $\beta=2$, $x=23$ and $153$). The
plots we show are representative of the evolution of the concentration
profiles along the downstream axis. The profiles are asymmetric with
respect to $Y=0$; in particular the maximum of the concentration is
not located at the channel center but is displaced towards the side of
the CaGreen solution. For increasing $X$, the reaction zone broadens.
The solid lines on Fig.~\ref{fig:experiment-numerics1} represent the
value of $c$ obtained by numerical solutions of Eqs.~\ref{eq:rd-nd}
and~\ref{eq:rc-nd}, by a standard one dimensional finite-difference
scheme with time stepping on a personal computer, using the values of
the parameters previously established. The offset on the right hand
side of the experimental curves is due to the background fluorescence
of the unbound CaGreen (reactant $a$). We match this offset in the
numerical curves by plotting $h a+c$ rather than just $c$, where $h$ a
normalization constant that gives the correct offset. This does not
affect the fitting parameters but matches the experimentally measured
quantity, i.e. the total fluorescence. We observe good agreement
between theory and experiment.

\bigskip

\begin{figure}[hb]
\includegraphics[width=9cm]{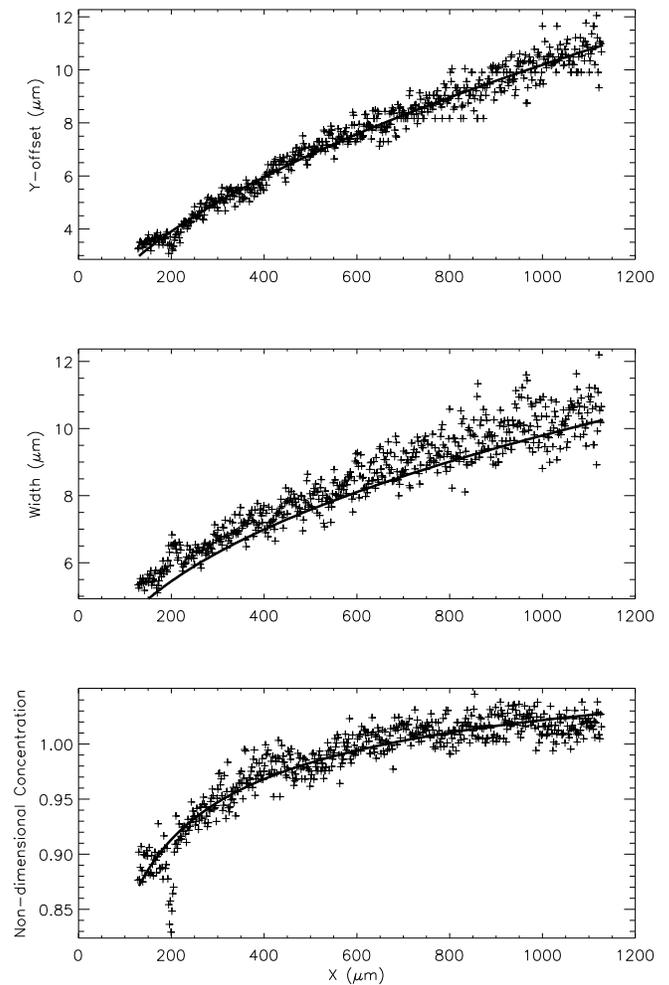}
\caption{Evolution of the position of the maximum concentration,
width, and value of the maximum with $X$. The $+$ signs are the
experiments, while the solid lines are the numerical
simulations. Here, $\beta=\sqrt{2}$,
$\ell=0.84$~\um, $U=0.334$~cm/s.\label{fig:position}}
\end{figure}

A quantitative measure of the agreement between the simulations and
experiments is obtained by looking at some representative quantities for
the profiles. In particular, we compare the downstream evolution of
three quantities: The position of the maximum, the width at half-height,
and the value of the maximum (Fig.~\ref{fig:position}). These evolutions
do not follow simple power laws in the full range of $x$ we consider.
However, for $x\gtrsim30$ dimensionless units, the power laws are better
defined. The useful scaling range in the experiments is further limited
(at $X\simeq1200$\um) by lateral wall effects, thus limiting the range
over which a power law can be observed; although the scaling range is
too small to obtain a clear exponent, our measurements are consistent
with the long time limit analytical theory of G\'alfi and
R\'acz~\cite{galfi88}. The maximum concentration shows a saturation
effect at large distances, i.e above $x\simeq300$. These measurements
can be compared with the numerics, plotted as full
lines. Fig~\ref{fig:position} shows there is good agreement between the
theory and the experiment. A similar agreement is obtained different
values of $\beta$ and flowrates $U$ we considered.

\bigskip
\bigskip

In conclusion, we demonstrate that two-dimensional reaction-diffusion
equations with stepwise initial conditions define not only a
theoretical problem, but also a physical situation which can be
achieved experimentally. Microfluidic reactors are particularly
convenient in order to produce such physical situations. Conversely,
2D reaction diffusion equations can be used to determine, from
experimental measurements, the relative concentration of the reactants
(micro-titration), the diffusion coefficients or the chemical
kinetics. In the present case, we demonstrate a determination of the
kinetic coefficient $k$ in order to match measured physical quantities
(position, width and concentration of the reactant) with numerical
predictions. The value of $k$ thus obtained for our system corresponds
to a reaction half-time of 1~ms, which is faster than the time scales
that can be resolved in a stopped-flow apparatus. This indicates that
the present approach can be used for analyzing fast chemical kinetics.

The authors acknowledge technical help from Bertrand Lambollez, Jean
Rossier and Thibault Colin, and useful discussions with Martin Bazant.


\begin{thebibliography}{13}
\bibitem[{rea()}]{reaction}
\bibinfo{note}{The pH of the solutions is stabilized at 7.5 by using a Tris
  buffer. Furthermore, traces of Calcium in the Ca-green solution are neutralized
  by adding excess EDTA.}
\end{thebibliography}

\end{document}